\documentclass[%
 reprint,
superscriptaddress,
preprintnumbers,
nofootinbib,
 amsmath,amssymb,
 aps,
 prb,
floatfix,
]{revtex4-2}

\usepackage{placeins}
\usepackage[utf8]{inputenc}
\usepackage[english]{babel}

\usepackage{amsfonts,amssymb,amsmath}
\usepackage{graphicx}
\usepackage{siunitx}
\usepackage{physics}
\AtBeginDocument{\RenewCommandCopy\qty\SI}
\ExplSyntaxOn
\msg_redirect_name:nnn { siunitx } { physics-pkg } { none }
\ExplSyntaxOff
\usepackage{multirow}
\usepackage{tabularx,booktabs}
\usepackage[dvipsnames]{xcolor}  % usenames is obsolete
\usepackage{soul}
\usepackage{dcolumn}
\usepackage{url}
\usepackage{fancyhdr}
\setcitestyle{numbers,square}
\usepackage{hyperref}

\begin{document}
\pagestyle{plain}

\title{%
  \texorpdfstring{%
    Electronic mean free path of the cuprate superconductor Bi$_2$Sr$_2$CaCu$_2$O$_{8+\delta}$ from thermal Hall conductivity%
  }{%
    Electronic mean free path of the cuprate superconductor Bi2Sr2CaCu2O8+δ from thermal Hall conductivity%
  }%
}

\author{Emma Campillo$^\S$}
\email{emma.campillo.munoz@usherbrooke.ca}
\affiliation{Institut quantique, D\'epartement de physique \& RQMP, Universit\'e de Sherbrooke, Sherbrooke, Qu\'ebec J1K 2R1, Canada}

\author{Manel Mezidi$^\S$}
\affiliation{Institut quantique, D\'epartement de physique \& RQMP, Universit\'e de Sherbrooke, Sherbrooke, Qu\'ebec J1K 2R1, Canada}
\affiliation{Universit\'e Paris–Cit\'e, Laboratoire Mat\'eriaux et Ph\'enomènes Quantiques, CNRS (UMR 7162), 75013 Paris, France}

\author{Lu Chen}
\affiliation{Institut quantique, D\'epartement de physique \& RQMP, Universit\'e de Sherbrooke, Sherbrooke, Qu\'ebec J1K 2R1, Canada}

\author{Ashvini Vallipuram}
\affiliation{Institut quantique, D\'epartement de physique \& RQMP, Universit\'e de Sherbrooke, Sherbrooke, Qu\'ebec J1K 2R1, Canada}

\author{Jordan Baglo}
\affiliation{Institut quantique, D\'epartement de physique \& RQMP, Universit\'e de Sherbrooke, Sherbrooke, Qu\'ebec J1K 2R1, Canada}

\author{Munkhtuguldur Altangerel}
\affiliation{Institut quantique, D\'epartement de physique \& RQMP, Universit\'e de Sherbrooke, Sherbrooke, Qu\'ebec J1K 2R1, Canada}

\author{Ga\"el Grissonnanche}
\affiliation{Laboratoire des Solides Irradiés, \'Ecole Polytechnique, 91128 Palaiseau, France}

\author{Genda Gu}
\affiliation{Condensed Matter Physics and Materials Science, Brookhaven National Laboratory (BNL), New York, USA}

\author{Louis~Taillefer}
\email{louis.taillefer@usherbrooke.ca}
\affiliation{Institut quantique, D\'epartement de physique \& RQMP, Universit\'e de Sherbrooke, Sherbrooke, Qu\'ebec J1K 2R1, Canada}
\affiliation{Canadian Institute for Advanced Research, Toronto, Ontario M5G 1M1, Canada}

\date{\today}

\begin{abstract}
We use thermal transport to access the electronic mean free path of $d$-wave quasiparticles in one of the most widely studied cuprate superconductors, Bi$_2$Sr$_2$CaCu$_2$O$_{8+\delta}$ (Bi2212). We have measured the thermal conductivity $\kappa_{\rm xx}$ and the thermal Hall conductivity $\kappa_{\rm xy}$ of three single crystals across a range of dopings. In the overdoped and optimally-doped samples, a clear enhancement is observed in both $\kappa_{\rm xx}$ and $\kappa_{\rm xy}$ upon cooling below the critical temperature $T_{\rm c}$, due to a suppression of the inelastic electron-electron scattering as electrons condense into pairs. The underdoped sample shows no enhancement in either, pointing to a high degree of disorder in that sample.
For the two highest dopings, the magnitude of the enhancement in $\kappa_{\rm xy}$ is controlled by the strength of the elastic impurity scattering. Using a prior model to estimate the mean free path from $\kappa_{\rm xy}$ data, we find that the mean free path in Bi2212 is approximately 7 times shorter than in YBa$_2$Cu$_3$O$_7$, considered to be one of the least disordered cuprates.
We conclude that the thermal Hall technique is a good way to compare the mean free path of $d$-wave quasiparticles in various cuprate materials.

\end{abstract}



\maketitle

\def\thefootnote{$\S$}\footnotetext{E.C.\ and M.M.\ contributed equally to this work.}\def\thefootnote{\arabic{footnote}}

\section{Introduction}

To this day, cuprate superconductors keep attracting significant attention, not only because they present the highest superconducting critical temperature, $T_{\rm c}$, at ambient pressure, but also due to the strong electronic correlations that give rise to a rich phase diagram. Cuprates are composed of CuO$_2$ planes separated by charge reservoir layers, which make them quasi-two-dimensional materials amenable to hole doping. As hole carriers are introduced, the system evolves from a Mott insulator to a Fermi liquid, with a range of quantum phases in between, including superconductivity.

When changing the carrier concentration through doping, this will introduce disorder in the crystal lattice, \textit{e.g.}, by doping with Sr in  La$_{2-x}$Sr$_x$CuO$_4$ (LSCO) or with O in YBa$_2$Cu$_3$O$_{6+x}$ (YBCO). This disorder will influence the electronic properties of the system. Because impurities have a strong effect on $d$-wave superconductors, there is a need to quantify the impact of disorder in cuprates.

Electrical resistivity is the most convenient technique for estimating the mean free path of electrons, since the zero-temperature value of the resistivity, $\rho_0$, is proportional to the impurity scattering rate. However, a measurement of $\rho_0$ is inaccessible in most cuprates, due to their high critical temperatures and correspondingly high upper critical fields. For materials like YBCO and Bi2212, suppressing superconductivity near optimal doping would require magnetic fields of the order of 150~T \cite{grissonnanche2014direct}, beyond what is achievable in the lab today.

The thermal Hall conductivity, $\kappa_{\rm xy}$, provides an interesting alternative to estimate the mean free path. Its advantange over electrical transport is that it can be measured \textit{inside} the superconducting state, at low fields. Furthermore, it is more suitable than the longitudinal thermal conductivity, $\kappa_{\rm xx}$, since phonons contribute significantly alongside electrons to the thermal conductivity: $\kappa_{\rm xx} = \kappa_{\rm qp} + \kappa_{\rm ph}$. Zhang \textit{et al.} were the first to use $\kappa_{\rm xy}$ to estimate $\ell_{\rm s}$, the mean free path in the superconducting state, an approach they applied to ultraclean YBCO at $p$ = 0.18  \cite{Zhang2001}. More recently, a similar approach was applied to HgBa$_2$CuO$_{6+\delta}$ (Hg1201) and HgBa$_2$Ca$_2$Cu$_3$O$_{8+\delta}$ (Hg1223) by Altangerel \textit{et al.} \cite{altangerel2025}.

Bi$_2$Sr$_2$CaCu$_2$O$_{8+\delta}$ (Bi2212) has been the cuprate for choice in many studies, not only for its accessible broad doping range but also because its quasi-2D nature makes it an ideal candidate for surface-sensitive techniques like ARPES \cite{Damascelli2003, Vishik2010, Vishik2012} and STM \cite{Pan2000, McElroy2005}. This has provided valuable information on several aspects, including direct evidence of intrinsic disorder \cite{Pan2000, McElroy2005}. It is important to quantify the level of disorder in Bi2212, and compare it to YBCO, the gold standard among cuprates. Indeed, YBCO is considered one of the cleanest cuprates, as evidenced by the observation of quantum oscillations \cite{DoironLeyraud2007} and the report of very large microwave conductivity \cite{Bonn1993}. In this work, we have measured the $\kappa_{\rm xy}$ of Bi2212 samples at three different dopings: underdoped, optimally doped, and overdoped. From our $\kappa_{\rm xy}$ data, and using the model previously developed by Altangerel \textit{et al.} \cite{altangerel2025}, we find that Bi2212 has an elastic mean free path, $\ell_{\rm s0}$, roughly 7 times shorter than YBCO.

\begin{figure}[t!]
\centering
\includegraphics[width=\columnwidth]{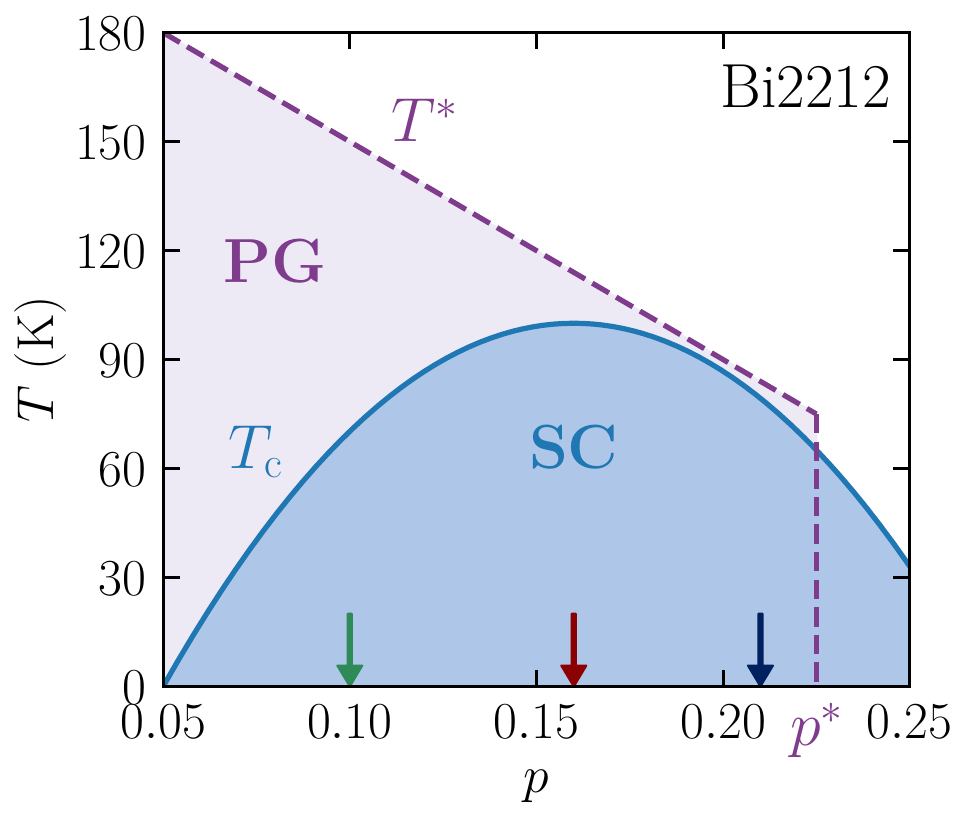}
\caption{Phase diagram of Bi2212 as a function of temperature and hole doping showing the superconducting dome (solid blue line) and the pseudogap phase (dashed purple line). The boundary of the pseudogap phase ($T^{\star}$) is taken from Ref. \cite{loret2017}. The hole concentration of the three samples in this study is marked by arrows: $p=0.10$ (green), 0.16 (red) and 0.21 (blue).}
\label{phase_diagram} 
\end{figure}

\section{Experimental Method}

\subsection{Samples}

We have measured thermal transport in three Bi2212 samples, at three different dopings across the phase diagram (Fig.~\ref{phase_diagram}). Single crystals were grown by the floating zone method \cite{wen2008} and their critical temperatures were obtained with a Physical Property Measurement System (PPMS) with Vibrating Sample Magnetometry (VSM) option. Their hole-doping levels were calculated using the expression of Presland and Tallon \cite{Presland_eq}, which assumes $p =$ 0.16 as optimal doping and that each oxygen ion corresponds to two holes. For the rest of the paper we will refer to each sample by its doping level: underdoped $p$ = 0.10 ($T_{\rm c}$ = 65 K), optimally doped $p$ = 0.16 ($T_{\rm c}$ = 91 K) and overdoped $p$ = 0.21 ($T_{\rm c}$ = 74 K).

\subsection{Technique}

The thermal transport measurement has been described in detail in Refs.~\cite{grissonnanche2019giant, chen2022large, altangerel2025}. We apply a heat current $\dot{Q}$ along the $a$ axis of the sample, and measure the temperature at two positions along the length of the sample, one closer to the heater, $T^{+}$, and a second one closer to the copper heat sink, $T^-$. These two temperatures were measured with two Cernox sensors for the $p =$ 0.10 and 0.21 samples, whereas for the $p = 0.16$ they were measured with a type-E thermocouple. The longitudinal thermal conductivity, $\kappa_{\rm xx}$, is defined as:
\begin{equation}
    \kappa_{\rm xx} = \frac{\dot{Q}}{\Delta T_{\rm x}}  \left( \frac{L}{wt} \right)
\end{equation}
%\vspace{0.5em}
where $L$ is the distance between $T^{+}$ and $T^{-}$, $w$ is the width of the sample, and $t$ is the thickness. The temperature difference along the $a$ axis is defined by $\Delta T_{\rm x} = T^+ - T^-$.

The thermal Hall effect was measured by applying a magnetic field, $B$, perpendicular to the CuO$_2$ planes, along the $c$ axis of the crystal. This creates a transverse temperature difference across the width, $\Delta T_{\rm y}$, measured using a differential type-E thermocouple for all three samples. This then defines the thermal Hall conductivity:
\begin{equation}
    \kappa_{\rm xy} = - \kappa_{\rm yy} \left( \frac{\Delta T_{\rm y}}{\Delta T_{\rm x}} \right) \left( \frac{L}{w} \right)
\end{equation}
where $\kappa_{\rm yy} = \kappa_{\rm xx}$ in Bi2212.
To eliminate any spurious contribution from the longitudinal thermal conductivity $\kappa_{\rm xx}$ due to a possible misalignment of the transverse contacts, we measure the transverse temperature difference $\Delta T_{\rm y}$ at both positive and negative magnetic fields, and extract the antisymmetric component $\Delta T_{\rm y}^{\rm as}(T,B)~=~[\Delta T_{\rm y}(T,B) - \Delta T_{\rm y}(T,-B)]/2$.

\section{Results and Discussion}

\begin{figure}[t!]
\centering
\includegraphics[width=\linewidth]{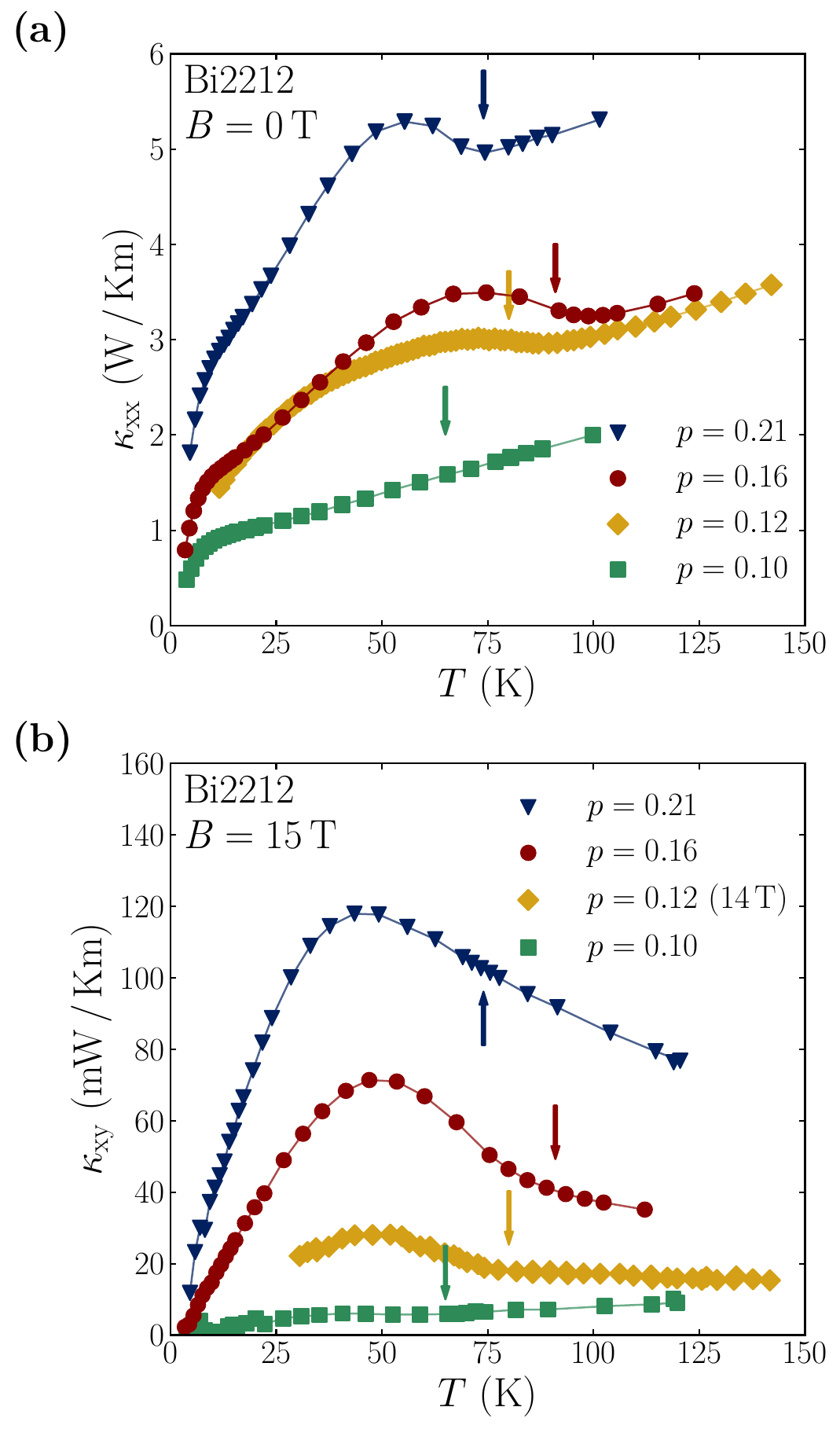}
\caption{Thermal conductivity $\kappa_{\rm xx}$ at $B = 0$ (a) and thermal Hall conductivity $\kappa_{\rm xy}$ at $B = 15$~T (b),
as a function of temperature for our three samples of Bi2212: $p = 0.10$ (green squares), $p = 0.16$ (red circles), and $p = 0.21$ (blue triangles). For comparison, we reproduce prior data taken on a Bi2212 sample with $p = 0.12$ (yellow diamonds) \cite{zeini2001}:
$\kappa_{\rm xx}$ at $B = 0$ and $\kappa_{\rm xy}$ at $B = 14$~T. The zero-field superconducting temperature $T_{\rm c}$ of each sample is marked by a color-coded arrow.}
\label{Kxx_0T}
\end{figure}

\begin{figure}[t!]
\centering
\includegraphics[width=\linewidth]{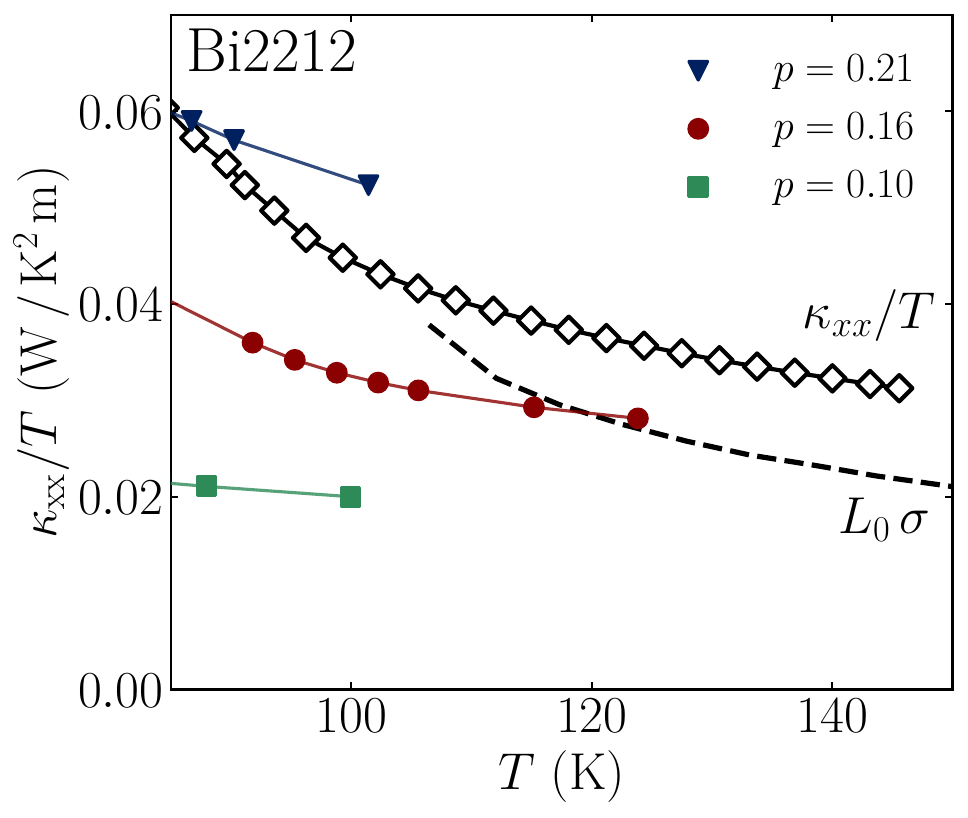} 
\caption{Thermal conductivity of Bi2212 at $B = 0$ for $T > 90$~K, plotted as $\kappa_{\rm xx}/T$ vs $T$ for our three samples: $p = 0.10$ (green squares), $p = 0.16$ (red circles), and $p = 0.21$ (blue triangles).
We reproduce prior data taken on an optimally-doped Bi2212 sample with $T_{\rm c} = 95$~K \cite{sun2008}, both for the $\kappa_{\rm xx}/T$ (open black diamonds) and for the electrical conductivity $\sigma$, plotted as $L_0 \sigma$ (black dashed line), 
where $L_0 = 2.44 \times 10^{-8}$~W~$\Omega$~/~K$^2$. This dashed line marks the upper bound on the electronic thermal conductivity, $\kappa_e/T$.}
\label{Ando_WF}
\end{figure}

\subsection{Thermal conductivity}

\begin{figure*}[!t] 
\centering 
\includegraphics[width=\linewidth]{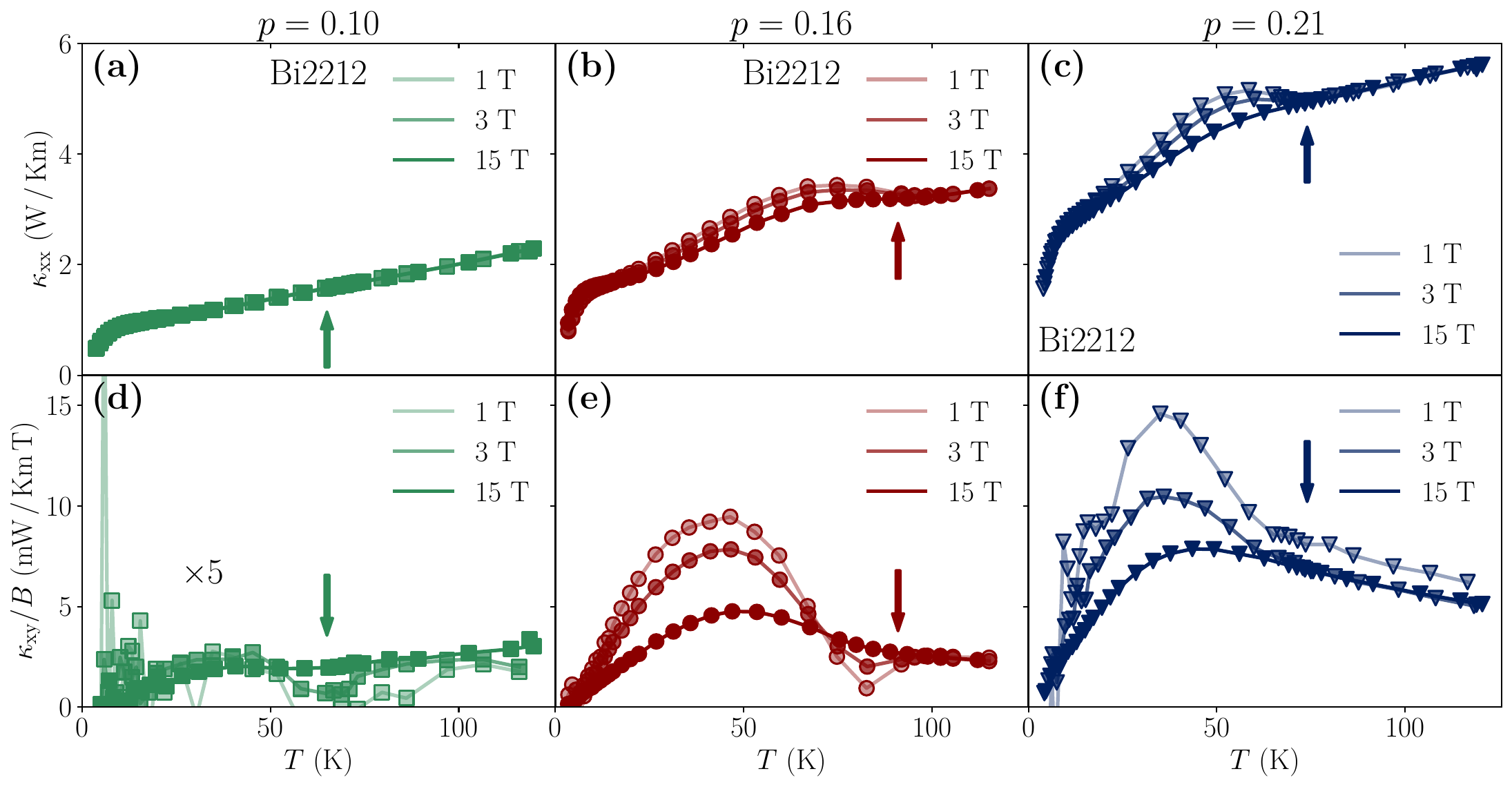} 
\caption{Longitudinal thermal conductivity $\kappa_{\rm xx}$ and thermal Hall conductivity divided by field $\kappa_{\rm xy}/B$ for Bi2212 as a function of temperature $T$ in magnetic fields of 1~T, 3~T and 15~T, with darker shades indicating higher fields. Lines serve as guides to the eye, while symbols represent the data points. Panels (a)–(c) show $\kappa_{\rm xx}$: (a) $p= 0.10$ (green); (b) $p= 0.16$ (red); (c) $p= 0.21$ (blue). 
Panels (d)–(f) show $\kappa_{\rm xy}/B$. Data in panel (d) is multiplied by 5. The arrows indicate the location of the superconducting transition.
} 
\label{Kxy_comparison} 
\end{figure*}

In Fig.~\ref{Kxx_0T}~(a), we present our thermal conductivity data vs temperature, in zero magnetic field. Our data agree reasonably well with previous data on Bi2212 at $p = 0.16$ \cite{zeini2001}.

In the normal state, above $T_{\rm c}$, we observe a monotonic increase of  $\kappa_{\rm xx}$ with doping, which we attribute to a growth in the electronic contribution. To validate this hypothesis, we reproduce in Fig.~\ref{Ando_WF} prior transport data on a Bi2212 sample at optimal doping \cite{sun2008}, comparing thermal conductivity (expressed as $\kappa_{xx}/T$) and charge conductivity (expressed as $L_0 \sigma$). We see that the conductivity due to electrons, $\kappa_e$, estimated via the Wiedemann-Franz law, $\kappa_{e}/T = L_0 \sigma$, is a sizable fraction of the total thermal conductivity. Knowing that the charge conductivity of Bi2212 increases rapidly with doping \cite{Watanabe1997}, as it does in all cuprates, it is reasonable to assume that $\kappa_{e}$ also does. Note, however, that the contribution from phonons, $\kappa_{\rm ph}$, still represents a significant fraction of $\kappa_{\rm xx}$ above $T_{\rm c}$, and it is therefore difficult to be quantitative about the relative contributions of electrons and phonons to $\kappa_{\rm xx}$.

Upon cooling below $T_{\rm c}$, we see from Fig.~\ref{Kxx_0T}~(a) that $\kappa_{\rm xx}$ exhibits a peak (below $T_{\rm c}$) at $p = 0.21$ and $p = 0.16$. The mechanism responsible for this peak is the following \cite{hirschfeld1996}: with decreasing temperature in the superconducting state, the system experiences a loss of quasiparticles as electrons condense into pairs, leading to a loss of heat carriers. However, the loss of quasiparticles also causes a decrease in the inelastic electron-electron scattering, which leads to an increase in the mean free path of each quasiparticle. As $\kappa_{\rm xx}$ is proportional to carrier density, these two opposing effects, a loss in inelastic scattering and an increase of mean free path, cause the peak in $\kappa_{\rm xx}$.

The height of the peak in $\kappa_{\rm xx}$ is controlled by the strength of inelastic scattering at $T_{\rm c}$ relative to the strength of elastic (impurity) scattering (at $T=0$). The absence of a peak in $\kappa_{\rm xx}$ for our sample with $p = 0.10$ (Fig.~\ref{Kxx_0T}~(a)) indicates that impurity scattering is comparatively much larger in that sample.

In other words, the peak in thermal conductivity contains information about the electronic mean free path. However, in order to be quantitative, one needs to avoid the contribution of phonons. This is why we turn to the thermal Hall conductivity, $\kappa_{\rm xy}$.

\subsection{Thermal Hall conductivity}

In Fig.~\ref{Kxx_0T}~(b), we plot the thermal Hall conductivity, $\kappa_{\rm xy}$, of our three samples, measured at $B = 15$~T, as a function of temperature. Again, our data agree reasonably well with previous data on Bi2212 \cite{zeini2001}. The magnitude of $\kappa_{\rm xy}$ above $T_{\rm c}$ is again seen to increase monotonically with doping. If $\kappa_{\rm xy}$ is predominantly caused by electrons, as we argue below, then this increase with doping confirms our interpretation (above) for the growth in $\kappa_{\rm xx}$ with doping.

The question is whether $\kappa_{\rm xy}$ is predominantly due to electrons. Before 2019, this was the general assumption since phonons were thought to generate only a very small thermal Hall effect \cite{Zhang2001,cvetkovic2015,zeini1999}. However, after 2019, it was realised that phonons also produce a sizable thermal Hall effect in cuprates \cite{grissonnanche2020chiral}. An important point is that the phonon contribution to $\kappa_{\rm xy}$ in cuprates is always negative \cite{grissonnanche2019giant,grissonnanche2020chiral,lizaire2021,boulanger2020,boulanger2022}. The fact that our $\kappa_{\rm xy}$ data on Bi2212 are positive at all dopings (Fig.~\ref{Kxx_0T}~(b)) shows that phonons are the minority carriers. To be more quantitative, let us compare our data with a typical phonon signal in cuprates. 

For example, in LSCO at a doping $p = 0.06$, where only phonons contribute, $\kappa_{\rm xy}$ = -~7~mW/Km at $T = 90$~K and $B = 15$~T \cite{grissonnanche2019giant}. For the same temperature and field, we find in Bi2212 that $\kappa_{\rm xy}$ = +~90~mW/Km at $p = 0.21$ and +~40~mW/Km at $p = 0.16$ (Fig.~\ref{Kxx_0T}~(b)). 
The positive electronic term is seen to overwhelm the typically negative phononic term. This means that the purely electronic component of $\kappa_{\rm xy}$ is likely to be only some 10-15\% larger than the measured value of $\kappa_{\rm xy}$.
Note, however, that $\kappa_{\rm xy}$ in our sample with $p = 0.10$ is much smaller (Fig.~\ref{Kxx_0T}~(b)), and thus we cannot neglect the phonon contribution in this case. We will not consider this sample in our analysis below.

In Fig.~\ref{Kxy_comparison}, we show our data for three different values of the magnetic field: 1, 3 and 15~T. In the lower panel, we plot the thermal Hall conductivity as $\kappa_{\rm xy}/B$ vs temperature. We see that the peak in $\kappa_{\rm xy}$ is most pronounced at the lowest field. This is simply because the field breaks pairs, and thus increases inelastic scattering. In the next section, we will use the data at $B = 1$~T to provide a quantitative estimate 
of the electronic mean free path in Bi2212.

Before doing so, we wish to point out two features of the data. First, we see in Fig.~\ref{Kxy_comparison}~(e) that there is a dip in $\kappa_{\rm xy}/B$ at $T_{\rm c}$, for the $p = 0.16$ sample at low fields
(1~T and 3~T). The origin of this intriguing dip, not seen in the overdoped Bi2212 sample (Fig.~\ref{Kxy_comparison}~(f)), remains unclear. Note however that a similar dip was observed in the cuprate Hg1223 at $p = 0.11$ (but not at $p = 0.09$ or 0.10) \cite{altangerel2025}.

Secondly, we see in Figs.~\ref{Kxy_comparison}~(d) and (e) that $\kappa_{\rm xy}/B$ is independent of field, \textit{i.e.} $\kappa_{\rm xy}$ is linear in field above $T_{\rm c}$. This behavior -- consistent with typical metallic behavior ~\cite{grissonnanche2019giant, grissonnanche2016} -- was also
observed in Hg1223, Hg1201 and YBCO \cite{altangerel2025}. However, in our Bi2212 with $p = 0.21$, we see that while the value of $\kappa_{\rm xy}/B$ above $T_{\rm c}$ is the same at 3~T and 15~T, it is not at 1~T. We believe this is an artifact of the 1~T data, possibly due to a movement of the thermal contacts after several cooldowns. This slight upward shift of the 1~T curve relative to the 3~T curve has minimal impact on our analysis of the mean free path (below), as the shift only exaggerates the estimated $\ell_{\rm s0}$ by about 10\%.

\subsection{Quasiparticle mean free path}

To obtain a quantitative measure of the disorder in our Bi2212 samples and compare it to that in other cuprates, such as YBCO, we estimate the electron mean free path using the model developed by Altangerel \textit{et al.} \cite{altangerel2025}, inspired by the work of Zhang \textit{et al.} \cite{Zhang2001}, which relies on a measurement of the thermal Hall conductivity $\kappa_{\rm xy}$ in the low-field limit (Eq.~8 in Ref.~\cite{altangerel2025}):

\begin{equation}
    \ell_{\rm s} = \ell_{B} \sqrt{\frac{2 \pi \hbar^2}{9 \zeta(3)k_{\rm B}^3}}\sqrt{\frac{v_{\Delta}k_{\rm F}d\kappa_{\rm xy}}{\eta T^2}},
\end{equation}
where $\ell_{\rm s}$ is the mean free path of nodal quasiparticles in the superconducting state, $\ell_B = \sqrt{\hbar/eB}$ is the magnetic length, $v_\Delta$ is the slope of the $d$-wave superconducting gap at the node, $k_{\rm F}$ is the nodal Fermi wavevector, and $d = 7.72$~\AA \ is the average interlayer separation between CuO$_2$ planes in Bi2212. The quantity $\eta$ is a dimensionless parameter characterizing the anisotropy of the scattering path length \cite{Zhang2001}; in the absence of detailed knowledge, we set $\eta = 1$ following Ref.~\cite{altangerel2025}.

From Eq.~(3), what we need is the value of $k_F$ and $v_\Delta$ in the nodal direction. ARPES studies of Bi2212 reveal that $v_\Delta$ is constant and equal to $1.5 \times 10^4$~m/s between $p=0.08$ and $p=0.20$~\cite{Vishik2010, Vishik2012}. It is well known that the nodal wavevector in Bi2212 varies only slightly between $p=0.10$ and $p=0.22$, with values close to $k_{\rm F} = 0.74$~\AA$^{-1}$~\cite{fujita2014}.

\begin{figure}[!t]
\centering 
\includegraphics[width=\columnwidth]{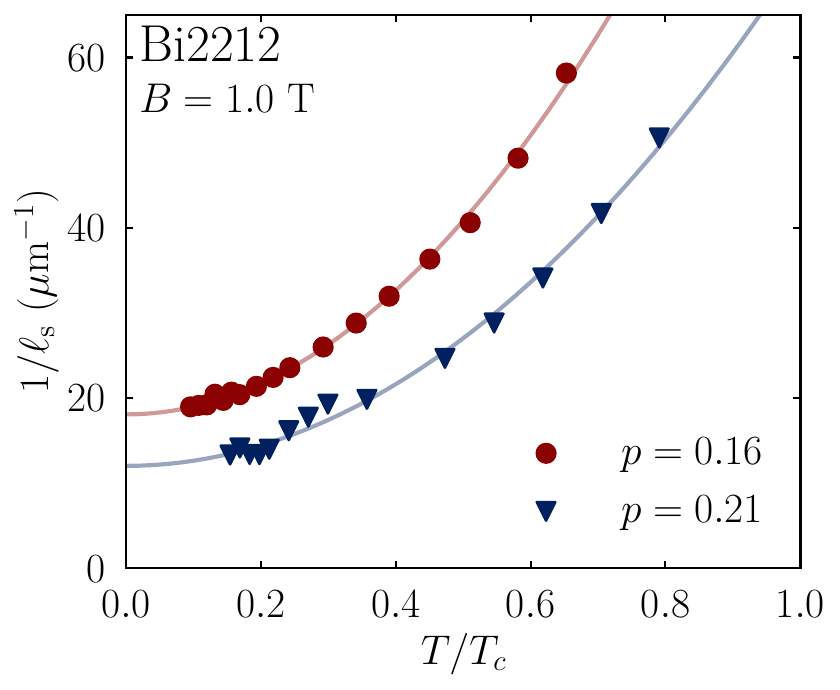} 
\caption{Inverse mean free path $1/\ell_{\rm s}$ as a function of reduced temperature for Bi2212 at $B = 1.0$~T: $p = 0.16$ (red) and $p = 0.21$ (blue).
Symbols represent the data points, while the lines correspond to quadratic fits ($a + b(T/T_{\rm c})^2$). 
The residual value at $T= 0$ is given by $1/\ell_{\rm s0} = 1/a$, whose values are listed in Table~I.
} 
\label{fitting_ell} 
\end{figure}

In Fig.~\ref{fitting_ell}, we plot the inverse mean free path, $1/\ell_{\rm s}$, which is proportional to the scattering rate, vs reduced temperature, for our Bi2212 samples with $p = 0.16$ and $p = 0.21$, where $\ell_{\rm s}$ is calculated using Eq.~(3) and the $\kappa_{\rm xy}$ data at $B = 1$~T in Figs.~\ref{Kxy_comparison}~(e) and \ref{Kxy_comparison}~(f). The data are well described by a quadratic fit, $1/\ell_{\rm s} = a + b(T/T_{\rm c})^2$, where the parameter $a$ represents the elastic scattering of nodal quasiparticles on impurities and the coefficient $b$ characterizes the strength of inelastic scattering.

In Fig.~\ref{fitting_ell}, such fits to our data allow us to extract the mean free path at $T = 0$, $\ell_{\rm s0}$, given by $\ell_{\rm s0} = 1/a$. For our two samples, we obtain $\ell_{\rm s0} = 550 \pm 120$~\AA, and $830 \pm 180$~\AA \ for $p= 0.16$ and 0.21, respectively. The error bars include the uncertainties on the fitting and on the geometric factor. In Table~I, we compare those two values for Bi2212 to the value obtained for an ultraclean sample of YBCO ($p = 0.18$)
using the same method \cite{altangerel2025}, as well as corresponding values for samples of the trilayer cuprate Hg1223 ($p = 0.10$) and
the single-layer cuprate Hg1201 ($p= 0.10$) \cite{altangerel2025}.

\begin{table}
\centering 
\begin{tabular}{|c| c |c |c |c |} 
\hline
Compounds & $T_{\rm c}$ (K) & $p$ & $d$ (\AA) & $\ell_{\rm s0}$ (\AA) \\ 
\hline 
\hline
Bi2212 & 91 & 0.16 & 7.72 & $550 \pm 120$ \\ 
Bi2212 & 74 & 0.21 & 7.72 & $830 \pm 180$ \\ 
YBCO & 90.5 & 0.18 & 5.8 & $3530 \pm 500$ \\
Hg1223 & 95 & 0.10 & 5.3 & $1590 \pm 230$\\
Hg1201 & 76 & 0.10 & 9.5 & $1040 \pm 150$\\
\hline 
\end{tabular} 
\caption{Properties of the measured samples, including the zero-field superconducting critical temperature $T_{\rm c}$, hole concentration (doping) $p$,
average inter-plane distance $d$, and mean free path $\ell_{\rm s0}$ in the $T=0$ limit for Bi2212 ($B = 1.0$~T) and for YBCO, Hg1223 and Hg1201 ($B = 0.5$~T) \cite{altangerel2025}. 
The error bars on $\ell_{\rm s0}$ include uncertainty from fitting and geometric factors.
} 
\label{table_mfp} 
\end{table}

\begin{figure}[t!]
\centering 
\includegraphics[width=\linewidth]{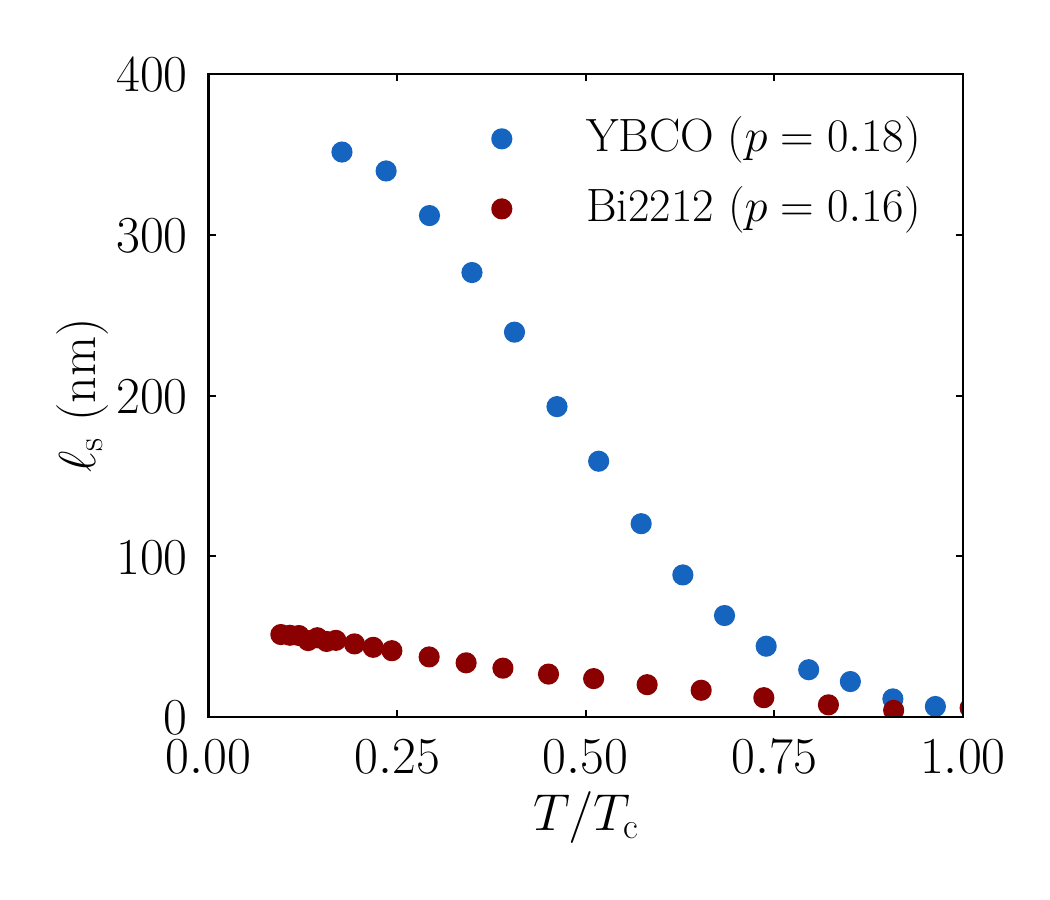} 
\caption{Mean free path of our Bi2212 sample with $T_{\rm c} = 91$~K ($p= 0.16$, red circles), 
as a function of reduced temperature ($T/T_{\rm c}$),
compared to published data on a YBCO sample with $T_{\rm c} = 90.5$~K ($p= 0.18$,
blue circles \cite{altangerel2025}).
In the $T = 0$ limit, $\ell_{\rm s0}$ is roughly 7 times smaller in Bi2212.
} 
\label{ell_OP91_YBCO} 
\end{figure}

In Fig.~\ref{ell_OP91_YBCO}, we plot the mean free path itself, $\ell_{\rm s}$ vs $T/T_{\rm c}$, for Bi2212 and YBCO with the same $T_{\rm c}$ value (91~K). We see that the impurity-controlled residual mean free path of YBCO is roughly 7 times larger than that of Bi2212.

Note that in principle the mean free path should be measured in the limit  $B \to 0$. Indeed, $\ell_{\rm s}$ obtained from Eq.~(3) depends on field, as found for clean samples of YBCO and Hg1223~\cite{altangerel2025}.  
However, in dirtier samples, like our Bi2212 samples, the field dependence is weak, and so comparing our estimate of $\ell_{\rm s0}$ obtained from data at $B = 1$~T is not very different from what it would be at $B = 0.5$~T, the field at which we plot $\ell_{\rm s}$ vs $T/T_{\rm c}$ for YBCO (Fig.~\ref{ell_OP91_YBCO}).

It is interesting to also make a comparison at lower doping. Our data on Bi2212 at $p = 0.10$ reveal a much smaller magnitude of $\kappa_{\rm xy}$ than at $p = 0.16$ or $p = 0.21$ (Fig.~\ref{Kxy_comparison}). This shows that the mean free path at $p = 0.10$ is much smaller than 500~\AA, although we cannot reliably quantify it. By contrast, in Hg1201 and Hg1223 at the same doping ($p = 0.10$), the values obtained are $\ell_{s0} = 1040 \pm 150$~\AA \ and $1590 \pm 230$~\AA \ (at $B = 0.5$~T) \cite{altangerel2025}. These comparisons based on $\kappa_{\rm xy}$ measurements, establish that the disorder in typical single crystals of Bi2212 causes a strong reduction of the electronic mean free path compared to YBCO, Hg1201 and Hg1223 (three of the cleanest cuprates, as manifested in the observation of quantum oscillations in all three \cite{DoironLeyraud2007,barivsic2013,oliviero2022}).

It is also of interest to consider the temperature dependence of the scattering rate, proportional to $1/\ell_{\rm s}$. Our data on Bi2212 show a quadratic $T^2$ dependence (Fig.~\ref{fitting_ell}). This contrasts with the cubic $T^3$ dependence observed in Hg1223 \cite{altangerel2025}, which is the dependence expected theoretically for a clean $d$-wave superconductor \cite{walker2000,dahm2005,duffy2001,quinlan1994}. We speculate that the departure from $T^3$ behavior seen in Bi2212 is caused by not being in the clean limit. It would be interesting to explore theoretically this dirty regime.

\section{Conclusion}

We have measured the thermal Hall conductivity of the high-$T_{\rm c}$ cuprate superconductor Bi2212, on single crystals with three different dopings: $p = 0.10, 0.16$ and 0.21.
From our data, we extract the quasiparticle mean free path in the superconducting state via a simple model. For our sample with $p = 0.16$, in which $T_{\rm c} = 91$~K, the residual mean free path at $T = 0$, controlled by disorder, is $\ell_{\rm s0} = 550$~\AA, compared to 3500~\AA \ in a YBCO sample with the same $T_{\rm c}$ ($p = 0.18$).
In our sample with $p = 0.10$, the thermal Hall conductivity is at least an order of magnitude smaller than in the samples of higher doping, such that the electronic mean free path in Bi2212 at $p = 0.10$ is much smaller than that of Hg1201 and Hg1223 at the same doping.
We conclude that the disorder level in typical samples of Bi2212 is such that the electronic mean free path of electrons in that material is much shorter than in the cuprates YBCO, Hg1201 and Hg1223.

We observe a quadratic temperature dependence of the inverse mean free path (or scattering rate), a clear departure from the $T^3$ dependence expected theoretically for a clean $d$-wave superconductor. We attribute this departure to the fact that Bi2212 is far from the clean limit. More work is needed to understand the $T^2$ behavior.

\section*{Acknowledgements}
We thank Seyed Amirreza Ataei and Étienne Lefrançois for fruitful discussions. We also thank Simon Fortier for his assistance with the experiments, and the cryogenics team at the Institut Quantique for their support. We are grateful to the Centre Tara for hosting a writing retreat that contributed to the completion of this manuscript. 
L.T. acknowledges support from the Canadian Institute for Advanced Research (CIFAR) as a CIFAR Fellow and funding from the Institut Quantique, the Natural Sciences and Engineering Research Council of Canada
(Grant No. PIN:123817), and the Canada Foundation for Innovation. 
G.G. acknowledges support from STeP2 No. ANR-22-EXES-0013, QuantExt No. ANR-23-CE30-0001-01, Audace CEA No. ANR-24-RRII-0004, and the École Polytechnique Foundation. 
E.C. acknowledges support from the Fonds de recherche du Québec through its Bourses d’excellence pour étudiants étrangers program \cite{FRQ351572}.
E. Campillo, M. Mezidi, A. Vallipuram, L. Chen, J. Baglo, M. Altangerel, and L. Taillefer have benefitted from their affiliation to the RQMP \cite{RQMP309032}. This research was undertaken thanks, in part, to funding from the Canada First Research Excellence Fund. This research project No. 324046 is made possible thanks to funding from the Fonds de recherche du Québec.
The work at BNL was supported by the US Department of Energy, office of Basic Energy Sciences, contract No. DOE-SC0012704.

\bibliography{reference}

\end{document}